\documentclass[final,5p,times,twocolumn]{elsarticle}
\usepackage{subfloat}

\usepackage{lineno}
\usepackage{xcolor}
\usepackage{placeins}
\usepackage[export]{adjustbox}

\journal{NIM-B}

\begin{document}

\begin{frontmatter}

\title{The TULIP project : first on-line result and near future}

\author[inst1]{V. Bosquet}
\author[inst1]{P. Jardin}

\author[inst2]{M. MacCormick}
\author[inst1]{C. Michel}

\affiliation[inst1]{organization={Grand Accélérateur National d'Ions Lourds, CEA/DRF-CNRS/IN2P3},%Department and Organization
            addressline={Bvd H. Becquerel}, 
            postcode={BP55027 14076 Caen cedex 5}, 
            country={France}}

\affiliation[inst2]{organization={IJCLab, Institut Joliot Curie Laboratory, CNRS/IN2P3},%Department and Organization
            addressline={15 Rue Georges Clemenceau},
            postcode={91400}, 
            country={France}}

\begin{abstract}
%% Text of abstract
The TULIP project aims to produce radioactive ion beams of short-lived neutron-deficient isotopes by using fusion-evaporation reactions in an optimized Target Ion Source System (TISS). The first step consisted of the design of a TISS to produce rubidium isotopes. It was tested with a primary beam of $^{22}$Ne@4.5 MeV/A irradiating a natural Ni target at the SPIRAL1/GANIL facility in March 2022. Rates of $^{76,78}$Rb were measured as well as an exceptionally short atom-to-ion transformation time for an ISOL system, of the order of 200 µs.
The second step of the project aims at producing neutron-deficient short-lived metallic isotopes in the region of $^{100}$Sn. A “cold” prototype has been realized to study the electron impact ionization in the TISS cavity and a “hot” version is under construction to prepare an on-line experiment expected in the near future.
\end{abstract}


\end{frontmatter}

\section{Introduction}
\label{sec:Introduction}
Since 2001, the SPIRAL1 (Système de Production d'Ions Radioactifs Accélérés en Ligne) facility at GANIL has been producing and accelerating RIBs (Radioactive Ion Beams). Initially, an electron cyclotron resonance ion source coupled with a thick carbon target produced mainly gaseous elements. In 2013 the facility was upgraded by replacing the ECR by a VADIS/FEBIAD ion source \cite{Penescu} and the same carbon target as before. This has allowed the range of available SPIRAL1 beam to be extended towards condensable elements \cite{Chauveau}.

A new TISS (Target and Ion Source Sytem) is in development for SPIRAL1 : The TULIP project (Target ion soUrces for short-Lived Ion Production)\cite{TULIP}. Its goal is the production of exotic neutron-deficient RIBs in the vicinity of $^{100}$Sn. The nuclear production mechanism is fusion-evaporation induced by the collision between the nuclei of an ion beam at an energy close to the Coulomb barrier energy and the atoms of a few µm thick metallic target. The TULIP project is divided into two steps, one for the production of rubidium and one for metals arround Sn. 

In exotic RIB production devices, the Atom to Ion Transformation time (AIT time, the average time between the production of the radioactive nucleus and their extraction as a RIB) competes with the radioactive decay. To limit the decay loss and so maximize the AIT efficiency, the AIT needs to be shorter than the half-life of the nucleus under production. The AIT is the sum of several times : diffusion out of the target or catcher material, effusion in the vacuum of the source, sticking of the atoms to the wall of the source and ionization and extraction of the ions. 

The goal of this project is to design a short-AIT TISS to produce short-lived isotopes with a good efficiency.

\section{Principle}

\begin{figure}[ht]
    \centering
    \includegraphics[max size={\columnwidth}]{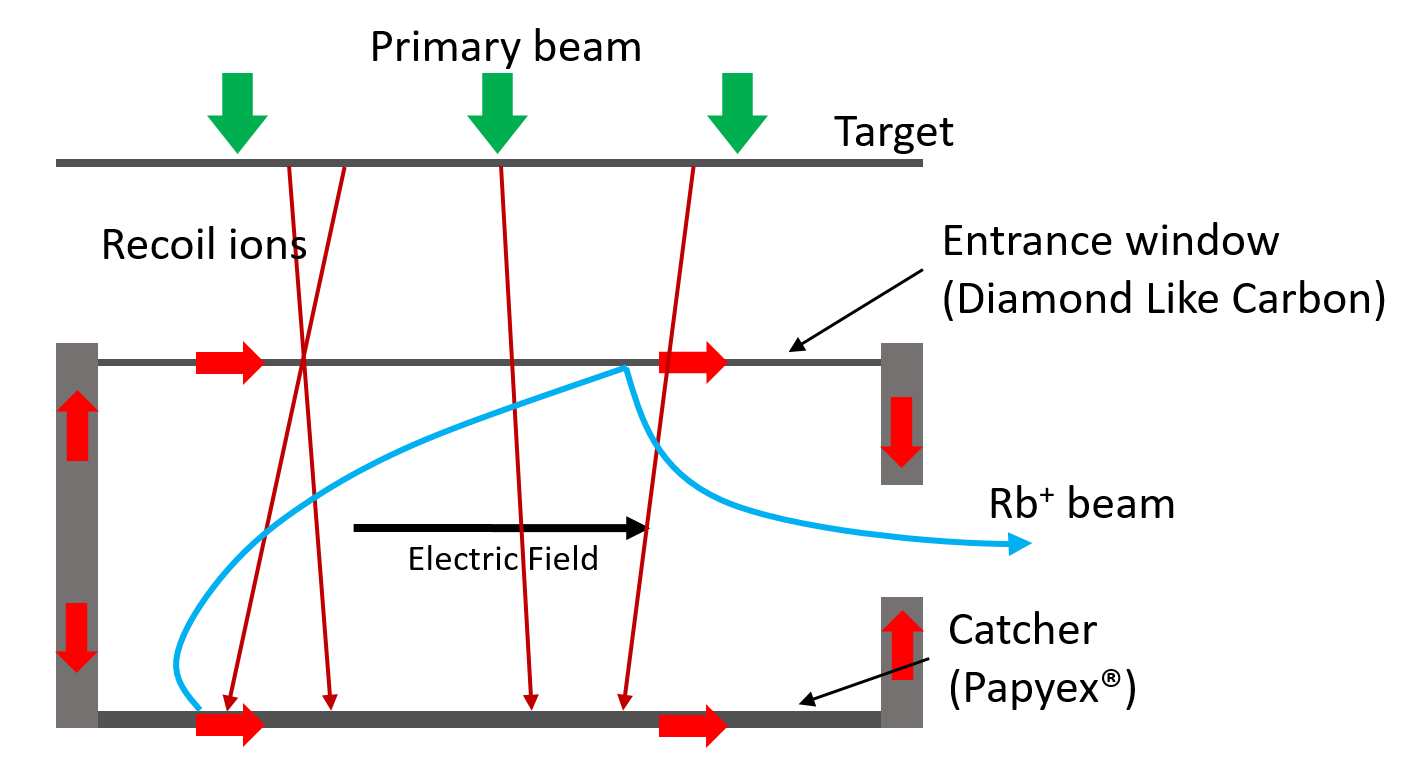}
    \caption{Principle of the surface ionization TISS.}
    \label{fig:SurfaceIonization}
\end{figure}

The surface ionization TISS consists of a 4~µm nickel target and a hot carbon cavity located 6 cm downstream from the target with respect to the primary beam direction (figure \ref{fig:SurfaceIonization}). The 4.5 MeV/A neon beam interacts with the target and produces radioactive rubidium among other nucleus that have enough energy to escape the target, enter the cavity and end up in a maximum depth of 10 µm in a graphite catcher. As the cavity is heated by a $\approx{300}$A current at 1300 °C, the implanted Rb diffuses in the catcher and reaches the surface in a short time. Due to its low first ionization energy the probability that the Rb leaves the C surface as an ion at this temperature is close to 100\%. The heating current generates an electric field in the cavity that push the Rb$^{+}$ ions toward the exit of the source. 

This TISS was tested on-line and produced $^{76-78}$Rb$^{+}$, with an AIT shorter than one millisecond [3]\cite{Jardin}. 

\begin{figure}[ht]
    \centering
    \includegraphics[max size={\columnwidth}]{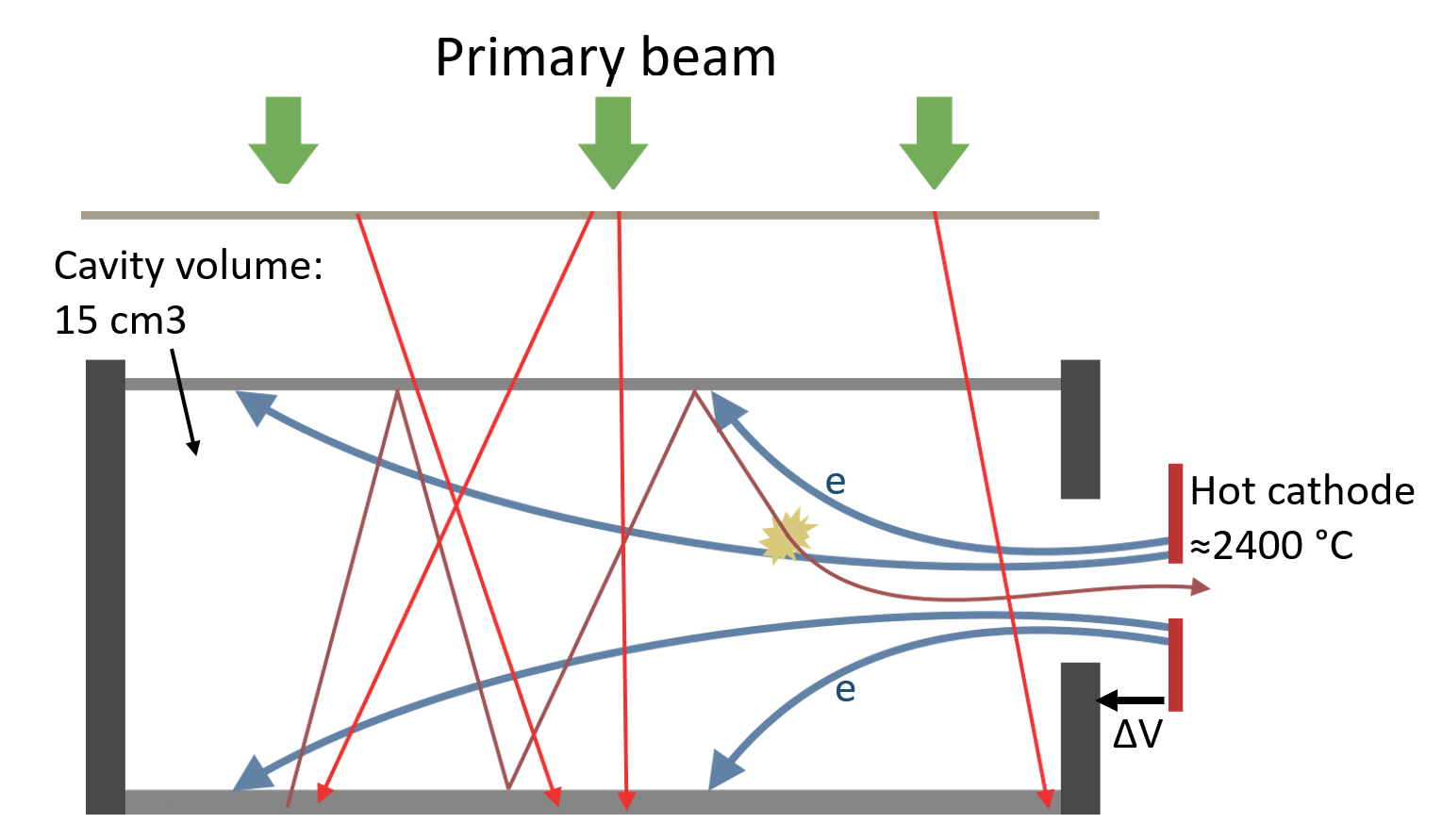}
    \caption{Principle of the electron impact ionization TISS.}
    \label{fig:ElectronImpactIonization}
\end{figure}

The electron impact ionization TISS (SPEED project : Système de Production d'Elements Exotiques Déficients en Neutrons) uses the same cavity with the addition of a hot tungsten cathode in front of the extraction hole. This cathode emits electrons that are accelerated and injected inside the cavity at 200 eV. The radioactive atoms effusing in the cavity are ionized by electron impact and extracted. The inner electric field results from three phenomena  : the voltage gradient along the source wall, the voltage between the cavity and the cathode and the electron density inside the cavity.

\section{Materials and Methods}

\begin{figure}[ht]
    \centering
    \includegraphics[max size=0.95\columnwidth]{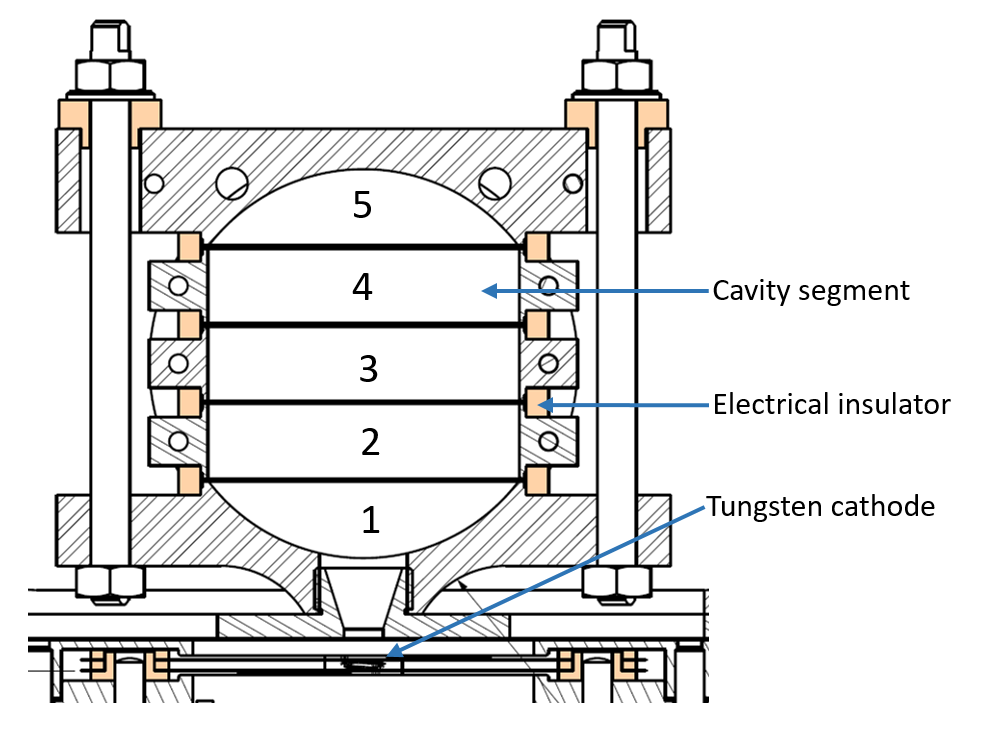}
    \caption{Cross section of the room temperature prototype. The cavity has a 40 mm diameter.}
    \label{fig:RoomTempPrototype}
\end{figure}

\begin{figure}[ht]
    \centering
    \includegraphics[max size=0.8\columnwidth]{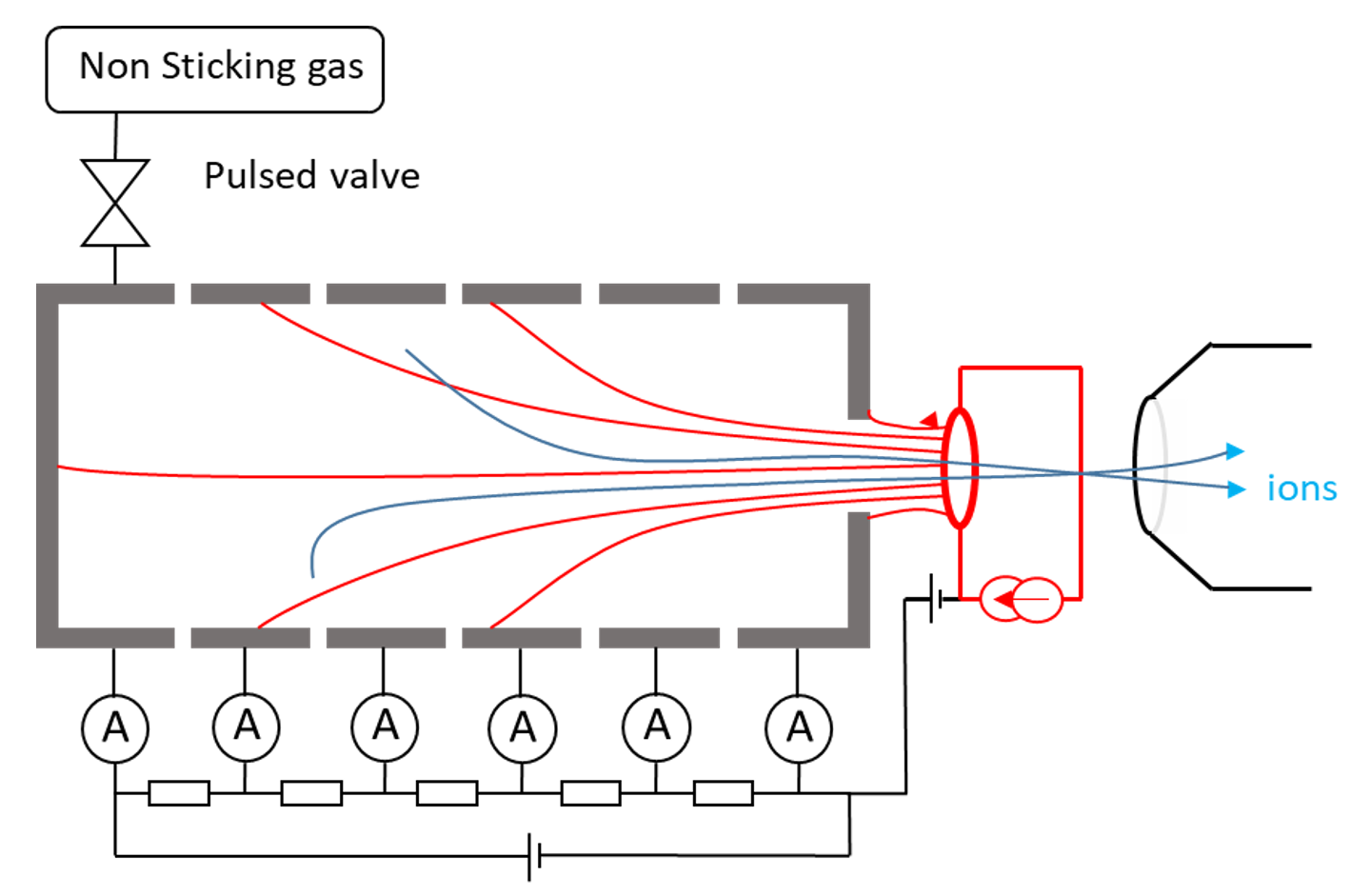}
    \caption{Diagram of the room temperature source.}
    \label{fig:principeSPEEDfroid}
\end{figure}

To allow the first proof of principle of the SPEED system to be performed on a laboratory bench, a prototype has been developed (figure \ref{fig:RoomTempPrototype}). The objective is to roughly reproduce the electric field present in the "hot" cavity ($\approx$1~V/cm). Therefore, the cavity has been divided in 5 five electrically insulated elements (fig. 3) biased by a voltage divider bridge. \ref{fig:principeSPEEDfroid}. The current collected on each segment is measured.

To measure the AIT time of the source, short pulses of noble gases are injected inside the source cavity through a quick valve. Because it is noble gases injected at room temperature, this response does not take into account the diffusion time nor the sticking time. The time necessary to extract an ion from within the cavity is assumed to be negligible, so the AIT measured takes into account only the atomic effusion. To extrapolate the AIT to other mass and temperature, corrections are applied. Indeed, in molecular flow mode, the average effusion time can be expressed as $\tau_{eff} \propto D_{eff} / v_{atoms}$ with $D_{eff}$ the average length traveled by atoms and $v_{atoms}$ their average speed. The length is constant whatever the effusing element, but the speed varies following the relation : $v_{atoms} \propto \sqrt{T / m}$ where T is the temperature in Kelvin and m the atomic mass. From this, the effusing time extrapolation formula is obtained : 

\begin{equation}
    \tau_{1} = \tau_{2} \sqrt{\frac{m_{1}T_{2}}{m_{2}T_{1}}}
    \label{equ:ExtrapolationTimeResponse}
\end{equation}

If the AIT is only dominated by atomic effusion and is experimentally measured at at known temperature, it is possible to extrapolate the AIT value for another effusing atom at an another temperature. 

\section{Experimental results}
\subsection{Electronic emission}

\begin{figure}[ht]
    \centering
    \includegraphics[max size=0.9\columnwidth]{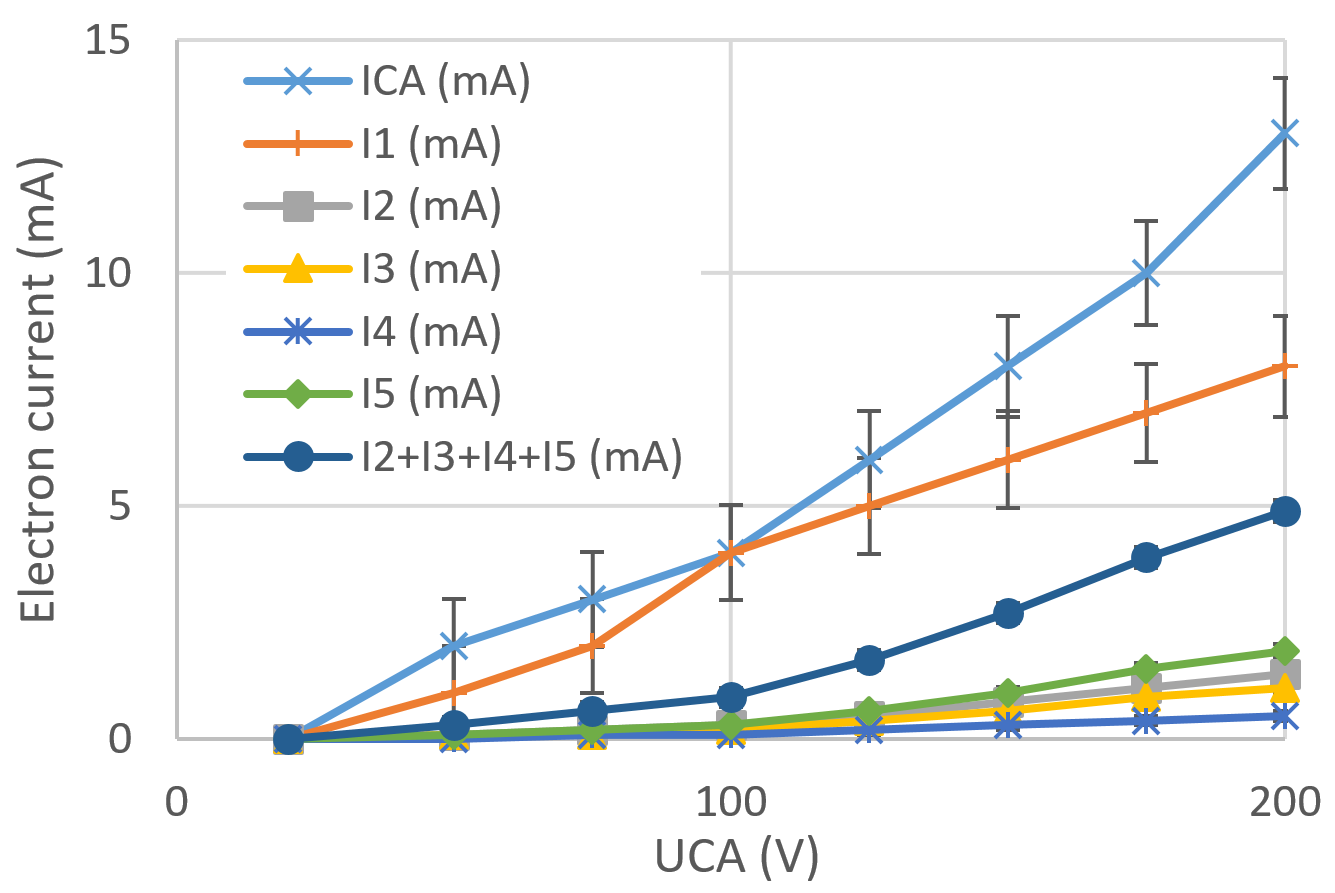}
    \caption{Current emitted by the filament (I\textsubscript{CA}) and current collected on each electrodes, as labelled on figure \ref{fig:RoomTempPrototype}.}
    \label{fig:Injected current}
\end{figure}

The tungsten filament cathode is set to a temperature of approximately 2400°C. Electrons emitted from the cathode are accelerated with a cathode-anode voltage U\textsubscript{CA} ranging from 0 to 200V. The current collected on each segment was measured versus U\textsubscript{CA}, as shown on figure \ref{fig:Injected current}.

The observation of an increasing current as a function of U\textsubscript{CA} indicates that the emission is space charge limited. In this regime, the current I\textsubscript{CA} emitted from the cathode approximately follows the Child-Langmuir law : $I_{CA} \propto {U_{CA}}^{3/2}$.

The electrons that participate the most in the ionization are the ones that penetrate inside the cavity : the sum of I\textsubscript{2} to I\textsubscript{5}. Since this sum continuously increase when UCA ranges from 0 to 200 V (as shown of figure \ref{fig:Injected current}), it is reasonable to assume that these currents could be greater for UCA larger than 200 V, thus increasing the ionization efficiency.

\subsection{Response time measurement}
\begin{figure}[ht]
    \centering
    \includegraphics[width=\columnwidth]{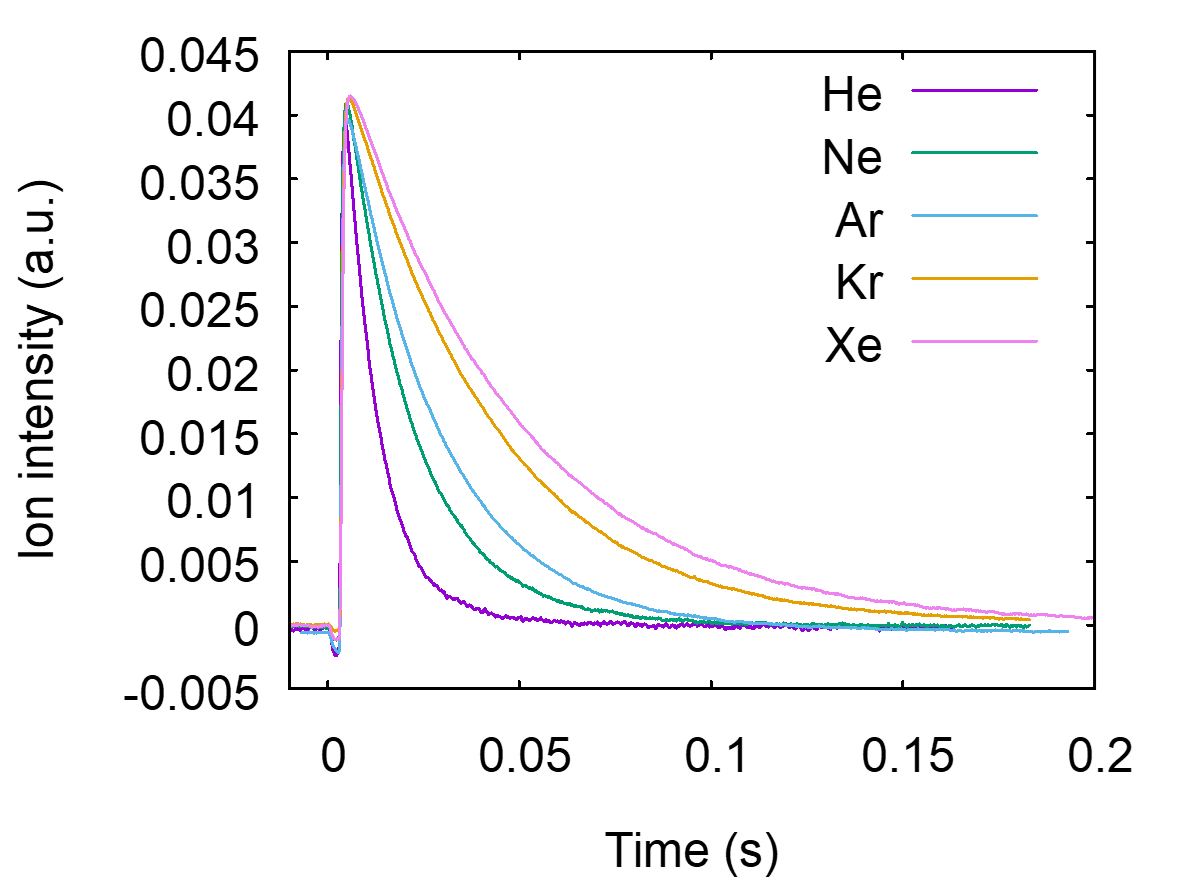}
    \caption{\label{first}Response of the source to pulses of nobles gases.}
    \label{fig:SignalHeXe}
\end{figure}

AIT has been measured with He, Ne, Ar, Kr and Xe. The valve opening command duration is 1 ms long. The ions exiting the source and cathode are accelerated by a 1 kV potential difference then focalised by electrostatic lenses. The ion current is then collected by a Faraday cup. Its intensity is measured as a function of time on an oscilloscope. The magnitude of the signal is the sum of a constant baseline created by the ionization of the residual gas and of the pulse of noble gas. Figure \ref{fig:SignalHeXe} shows the signal for each noble gas. The baseline is substracted and the magnitude normalised for ease of comparison.

\begin{figure}[ht]
    \centering
    \includegraphics[width=0.8\columnwidth]{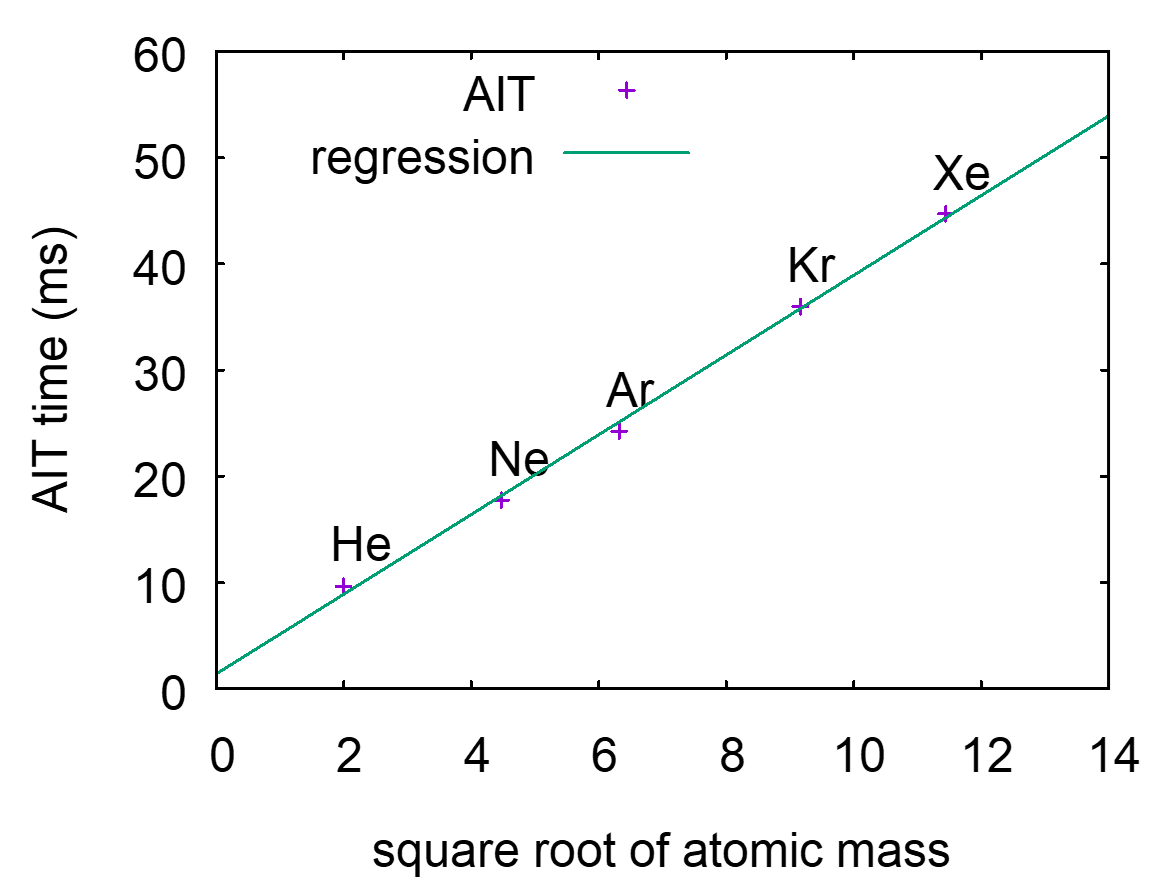}
    \caption{\label{second}AIT of the room temperature source for each noble gases.}
    \label{fig:TAIHeXe}
\end{figure}

When the valve opens, the ion current reaches a maximum after ~2 ms, corresponding to the time needed to reach an equilibrium between the invasion of the cavity by the gas and the release of atoms and ions through the exit hole. Then the ion intensity exponentially decays, with a decay time governed by the disappearing process of the atoms and ions from the cavity. For each noble gas, the decay has been fitted with the following function : $I_{ion}(t) = K exp(-t/AIT)$ where AIT is the average time taken by a noble gas atom to leave the source. The resulting AIT parameter is plotted as a function of root of the mass of the injected gas. Results in figure \ref{fig:TAIHeXe} shows the clear proportionnality between AIT and $\sqrt{A}$ following equation \ref{equ:ExtrapolationTimeResponse} and that the AIT is dominated by atomic effusion at room temperature. The polarization of the cavity has no measured influence on the AIT.

\subsection{Ionization efficiency and effect of the cavity polarization}
The source ionization efficiency has been measured for Ar. The pulsed valve was replaced by a calibrated leak, so that a known flow of Ar was injected into the source. The measurement of the Ar+ ion current allows to determine the AIT efficiency. The Ar+ ions were mass over q separated by using a magnetic dipole. A maximum AIT efficiency of 2\% was obtained, using a maximum U\textsubscript{CA} voltage of 270 V and a cavity polarisation of 5 V. 

The efficiency is maximized when the cavity polarization pushes the ions toward the exit. The efficiency is divided by two if the polarization is turned in the opposite direction and by 1.5 with no polarisation. The polarization has no perceivable effect on the electron current injected into the source, and no modification of I\textsubscript{CA}.

\section{Conclusion}
 The room temperature prototype of SPEED has been tested for the first time. A current of 5 mA was injected in the cavity and an AIT efficiency of 2\% has been obtained. The polarization of the cavity has shown its importance on the AIT efficiency. An increase of the electron current can certainly be obtained by playing on the mechanical configuration and on the voltage applied between the anode and the cathode. We thus expect to improve the present AIT efficiency. 

The AIT time has been measured for several noble gases at room temperature. One can deduce that for 100Sn, the effusion time at a temperature of 1400°C, without sticking, would be equal to 16 ms.

The AIT efficiency for Ar was measured. A similar value can be expected for an atom with an atomic mass around A=100 at 1400°C since the ionization efficiency is proportional to the product of the effusion time and the ionization cross section, and both are similar to Ar at 25°C. Further tests and modifications will be performed with the aim of increasing the ionization efficiency.

\end{document}